# S-matrix Fluctuations

# in a model with Classical Diffusion

# and Quantum Localization

Fausto BORGONOVI [a,c] and Italo GUARNERI [b,c]

[a] *Dipartimento di Fisica Nucleare e Teorica, Università di Pavia,*

*Via Bassi 6, 27100 Pavia, Italy*

[b] *Università di Milano, Sede di Como, Via Castelnuovo 7, 22100 Como, Italy*

[c] *I.N.F.N, Sezione di Pavia, Via Bassi 6, 27100 Pavia, Italy*

**Abstract:**

The statistics of S-matrix fluctuations are numerically investigated on a model for irregular quantum scattering in which a classical chaotic diffusion takes place within the interaction region. Agreement with various random-matrix theoretic predictions is discussed in the various regimes ( ballistic, diffusive, localized )

Erratic fluctuations of scattering cross sections in dependence of energy or other control parameters are a well known phenomenon in several branches of quantum physics. Stochastic methods which model the Scattering Matrix by a random matrix chosen in an appropriate ensemble have proven very successful in analyzing the statistics of these fluctuations. These methods, originally started in Statistical Nuclear Physics, are now widespread in other fields of microphysics[1]; of particular importance among these, the analysis of conductance fluctuations of small metallic samples at low temperature. A possible unifying approach to such disparate fluctuation phenomena has emerged from studies on the quantum dynamics of classically chaotic systems. Indeed, recent developments in that field indicate that cross section fluctuations distinguished by some universal features generically appear in quantum scattering processes when the corresponding classical dynamics is chaotic inside the interaction region[2, 3]. In some model systems exhibiting classical chaotic scattering, the statistics of the rapid S-matrix fluctuations occurring on energy scales comparable to the mean level spacing have been found to closely agree with the predictions of Random Matrix Theory, and specifically with those of Dyson's Circular Orthogonal Ensemble (COE) in the case of time reversal invariant scattering processes. In other cases, more specific matrix ensembles have been constructed in order to account for the presence of fast direct processes besides the chaotic one, and again a good agreement has been found [4, 5]. These results suggest that classical chaos may eventually provide dynamical grounds for the use of Random Matrix theoretical methods.

Analyzing conductance fluctuations from the standpoint of quantum chaotic scattering is a promising task. Some significant results in this direction have been obtained for the ballistic regime[6], but to the best of our knowledge the fluctuations of the S-matrix have not yet been investigated in situations exhibiting at once classical chaotic diffusion and quantum localization. This is the subject of the present Letter, where this issue is investigated on a recently introduced one dimensional model that can be analyzed both classically and quantum mechanically[7, 8]. The main qualitative features of this model are as follows. In the classical case it displays a chaotic diffusion leading to ohmic conductance. In the quantum case, the interplay of diffusion and localization determines a transition from a ballistic regime, where the sample length is comparable to the mean free path, to a localized regime, in which the sample length is larger than the localization length. At intermediate lengths between these two regimes, the quantum transmission coefficient exhibits an ohmic dependence on the sample length. This crossover region increases in size on moving towards the classical limit, and within it the magnitude of transmission fluctuations is scarcely dependent on the sample length. Thus the model exhibits some essential features of classical and quantum transport, and the analysis of its scattering fluctuations is relevant for the above sketched general themes. In this Letter we compare the numerically computed statistics of the S-matrix with the predictions of random matrix ensembles.

We shall now briefly describe the model, which was presented in detail in refs.[7, 8]. It describes the discrete time ( stroboscopic ) dynamics of a particle on a line. The scattering region ( in solid state language, the "sample" ) is an interval of length $L$. The one-step classical evolution is defined by:

$$\overline{n} = n - V'(\theta) \qquad \overline{\theta} = \theta + \tau\overline{n} \quad (mod\,(2\pi)). \tag{1}$$

where $n$ defines the position of the particle on the line, $V(\theta) = k\cos\theta$ with $k$ a fixed parameter, $\tau$ is zero outside the scattering region defined by $\mathcal{M} : n_0 \leq n < L + n_0$ and has a constant nonzero value ( also denoted by $\tau$ ) inside the region itself. The interacting dynamics is thus defined by the Standard Map; the free dynamics is just uniform motion at a constant speed $k\sin\theta$. Changes of the phase $\theta$ occur only inside the sample and can be thought of as "collisions". The mean distance travelled between subsequent collisions is on the order of $k$ which therefore defines the mean free path (mfp). In the deeply chaotic regime of the Standard Map ($k\tau >> 1$) chaotic transport takes place inside the sample, and the transmission across the sample scales as $L^{-1}$.

In the quantized model (with $\hbar = 1$) $n$ takes integer values and the one-step evolution is specified by the unitary propagator:

$$\mathcal{U} \equiv \mathcal{T}U_0 = [\sum_{n \in \mathcal{M}} e^{-in^2\tau/2}|n\rangle\langle n| + \sum_{n \in \mathbf{Z}\setminus\mathcal{M}} |n\rangle\langle n|]e^{-iV(\theta)}. \tag{2}$$

The interacting dynamics defined by (2) and the free dynamics given by the propagator $U_0 = e^{-iV(\theta)}$ give rise to a well–posed scattering problem : the Møller wave operators and the Scattering operator are mathematically well defined and so is the Scattering Matrix $S_{\alpha,\beta}(\lambda)$ at fixed quasi–energy $\lambda$. The scattering channels labelled by $\alpha$ are defined by free waves with wavenumbers $\theta_\alpha$ satisfying $k\cos\theta = \lambda \quad (mod\,2\pi)$. For given $\lambda$ there are $\sim 2k/\pi$ channels, and associated with each channel $\theta_\alpha$ there is a reversed channel $\theta_{\alpha*}$ that corresponds to a wave propagating in the opposite direction. The Scattering Matrix is obtained in the standard way from distorted plane waves $u_+^{\lambda,\alpha}$ the latter are in turn solutions of the Lippman–Schwinger equation

$$[1 - e^{i\lambda}(U_0 - e^{i\lambda+0^+})^{-1}(\mathcal{T}^\dagger - 1)]u_+^{\lambda,\alpha} = u_0^{\lambda,\alpha} \tag{3}$$

where $u_0^{\lambda,\alpha}$ are free waves. The Scattering matrix $S_{\alpha\beta}(\lambda)$ for a given quasi-energy $\lambda$ can be numerically computed with a great accuracy (unitarity error $\sim 10^{-6}$ for sample lengths up to $\sim 600$). In our numerical computations we actually take $V(\theta) = 2q\arctan(\xi\cos\theta)$ that becomes $k\cos\theta$ in the limit $q \to \infty, \xi \to 0, 2q\xi = k$. The rank of the scattering matrix is $\sim 2k/\pi$ and becomes large in the classical limit ($k \to \infty, \tau \to 0, k\tau = const., L/k = const.$).

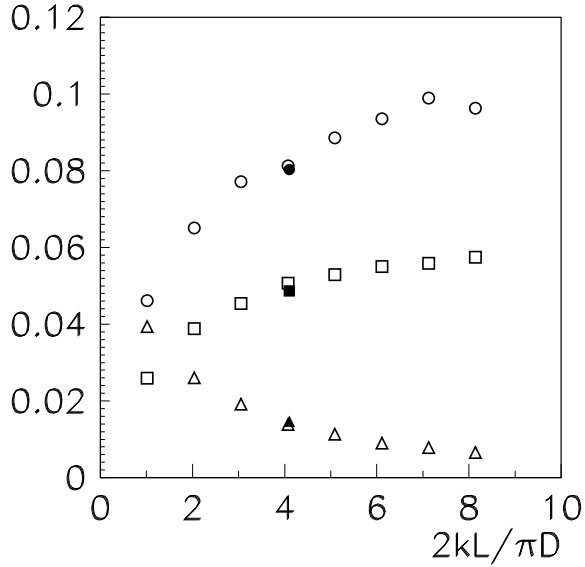

Figure 1. *Averaged squared moduli of S-matrix elements versus rescaled sample length from an ensemble of 100 matrices of rank $N = 30$, with $k\tau = 10$, $k = 45$, $\ell = 880$, $30 < L < 240$. Open circles: diagonal reflection matrix elements. Open squares: nondiagonal reflection elements. Open triangles: transmission elements. Full symbols at scaled length $\approx 4$ are obtained by averaging over a quasi-energy interval.*

Characteristic length scales are given by the mfp $\sim k$ and by the localization length $\ell$. The latter is semiclassically on the order of the classical diffusion coefficient $D$, which is in turn proportional to $k^2$. The mfp is the one classically relevant scale and the classical transmission coefficient depends on the *scaled sample length* $\mathcal{L} = 2kL/\pi D \sim L/k$, that is, on length measured in units of the mfp. The ballistic regime is defined by $\mathcal{L} \sim 1$ and the classical diffusive regime by $\mathcal{L} >> 1$, meaning that a large number of collisions occurs inside the sample. In addition to this, the quantum diffusive regime requires $L < \ell$, and these two conditions can be met the more easily, the larger $k$.

An ensemble of S-matrices is generated by shifting the sample along the $n-$axis. Under a well-known assimilation of the quantum standard map to a solid-state model of the tight-binding type, moving the sample to different positions is equivalent to taking different realizations of the disorder, while keeping the sample in a fixed position. In other models for chaotic scattering there was no "disorder averaging" and the statistics were generated instead by slightly changing the value of the energy. We have made a systematic use of ensemble averaging, but in a few test cases averages over small quasi-energy intervals were also computed and a close agreement with ensemble averages was found (compare for example the full symbols with the open ones in Fig.1).

The S-matrix has one symmetry expressed by the reciprocity relation $S_{\alpha\beta} = S_{\beta^*\alpha^*}$ Thanks to this property the matrix $\tilde{S}_{\alpha\beta} = S_{\alpha\beta^*}$ is unitary and symmetric; this very matrix was the object of our statistical analysis. In the rest of this Letter it will be denoted just $S$ and no further reference to the original matrix will be made. In Ref.[7] one important prediction of the theory of irregular scattering was tested, concerning the correlation of S-matrix elements at different quasi-energies spaced by $\epsilon$; it was found that in the ballistic regime the squared modulus of such correlations depends on $\epsilon$ in an approximately Lorentzian way, as expected on semiclassical grounds[2].

In this paper we focus on the statistics of S-matrix at fixed quasi–energy. Fig. 1 shows the average of squared moduli of three different classes of matrix elements : (a) diagonal elements (associated with reflection with unchanged velocity : open circles), (b) non diagonal elements (open squares) and (c) transmission elements (open triangles). For each class the recorded average was obtained by a double averaging process, first over all the matrix elements in the given class, and then over the matrix ensemble. The results are shown for values of the scaled length that, as shown in ref[7], correspond to the ballistic and the quantum ohmic regimes.

COE is certainly not a proper reference ensemble for the diffusive regime, because transmission and reflection elements have systematically different magnitudes and thus violate the COE invariance properties. On the other hand, even in the ballistic regime fast direct reactions are present, connected with classical orbits that do not undergo significant randomization inside the interaction region. These direct reactions give a non–fluctuating contribution to the S-matrix and have to be filtered out somehow.

To this end we used the quite general method described in ref.[5]: first we computed an ensemble–averaged (nonunitary) scattering matrix $\overline{S} = \langle S \rangle$ that could be turned into a diagonal matrix $\overline{S}'$ by a suitable transformation, ($\overline{S}' = U\overline{S}U^T$, with $U$ unitary and $U^T$ its transposed matrix). On applying the same transformation to the original matrix $S$ a new symmetric unitary matrix $S'$ was obtained, which is diagonal "on the average". Finally, we investigated the statistics of the "fluctuation" matrix $S^{fl} = S' - \overline{S}'$, searching for universal characters that could be modelled by Random Matrix Theory. In particular, we have compared our numerical results with the prediction of a general random matrix model which was found to satisfactorily reproduce the fluctuations observed in a dynamical model[5]. In the approach of Ref.[5] the average S-matrix is used as an input information in the building of a theoretical ensemble. The latter is uniquely defined by the coefficients $T_\alpha = 1 - |\overline{S}'_{\alpha\alpha}|^2$; in particular, given the $T_\alpha$'s, theoretical values for the 2nd moments $< |S^{fl}_{\alpha\beta}|^2 >$ can be found by numerically computing certain multifold integrals [9]. In a number of ballistic cases ($\mathcal{L} \approx 1$) we found the empirical distributions of $S^{fl}_{\alpha\beta}$ to be Gaussian with good accuracy. Empirical values of $< |S^{fl}_{\alpha\beta}|^2 >$ from an ensemble of 100 matrices were found to agree with theoretical ones within an average 20% accuracy, both for diagonal and for off–diagonal elements. A comparison for diagonal elements is shown in Fig.2. Although there is some qualitative agreement (in the non–diagonal case, the theoretical and the empirical 2nd moments remain within 20% of each other while changing over 15 orders of magnitude) the observed deviation is significantly larger than statistically expected; also, Fig.2 gives a clear indication for the theoretical values systematically underestimating the empirical ones. In the diffusive regime the agreement becomes worse and worse, the discrepancies being particularly large at small values of the coefficients $T_\alpha$, where theoretical data decrease and empirical ones increase instead. However in the diffusive case the random model is not even theoretically adequate (it assumes all the bound states to be equally coupled, in a statistical sense).

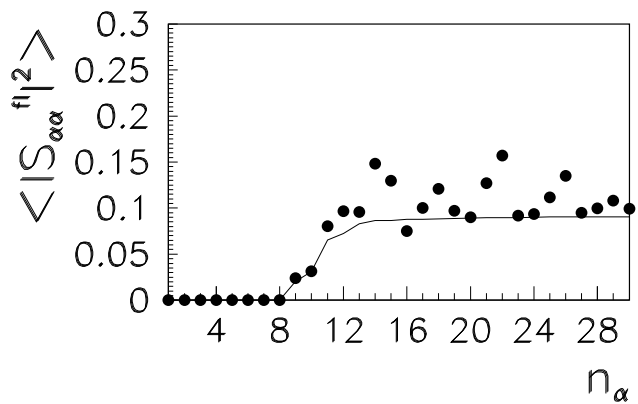

**Figure 2.** *Variance of diagonal elements of the fluctuating part of the $S$—matrix, for the same data as Fig.1 and $L=28$. The labeling of matrix elements given on the horizontal axis is taken so that the coefficients $T_\alpha$ monotonically increase from left to right. In order to facilitate comparison, a continuous line has been drawn through the theoretical data.*

If the scaled length is increased still further, the localized regime is entered, where transmission elements become exponentially small. The statistics of the transmission coefficient $T$[12] in the localized regime presents some elements of interest. According to our data, not reported here, the distribution of $T$ in this regime is approximately lognormal, consistently with theoretical predictions for quasi - 1d Anderson insulators.

In the far localized regime the S-matrix of rank 2N is approximately decomposed in two submatrices of rank N describing reflection on the left and on the right of the sample. In the limit of an infinitely long sample, these matrices become unitary and symmetric themselves. In this regime we have found the statistics of the reflection submatrices to closely agree with COE predictions. In Fig.3 the distributions of squared moduli of diagonal (a) and nondiagonal (b) reflection matrix elements are shown for a localized case $\ell/L \approx 1/8$, together with the theoretical distribution derived in ref.[10] (the latter is not an exponential one because of the small rank of the matrix, and in the nondiagonal case it involves an hypergeometric function). The average squared modulus of non diagonal elements is $\approx 1/(N+1)$ and that of diagonal elements is $\approx 2/(N+1)$. The factor 2 accounts for backscattering[11]. Here again the distributions were generated by the double averaging process described above. An important remark is that this agreement is here found in a highly nonclassical regime, dominated by the purely quantal localization effect. Indeed, one important prediction about S-matrix fluctuations in chaotic scattering, that was based on semiclassical considerations[2], is here violated: the squared-modulus of the correlation of $S$−matrix elements at *different* quasi-energies is not of Lorentzian type. In Fig.4 we show a typical correlation,

$$C_{\alpha\beta}(\epsilon) = \frac{|\langle S^*_{\alpha\beta}(0) S_{\alpha\beta}(\epsilon) \rangle|^2}{\langle |S_{\alpha\beta}(0)| \rangle^2} \quad (4)$$

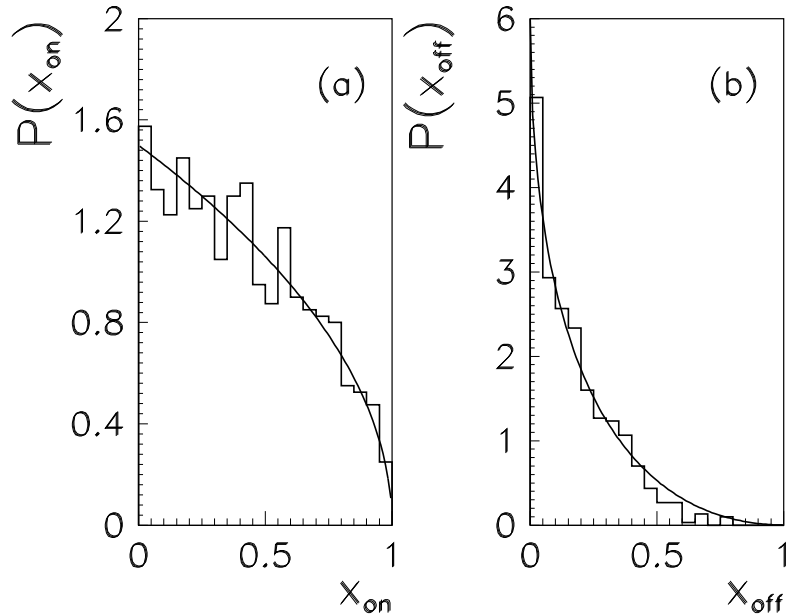

Figure 3. *Normalized distributions of the squared moduli of the S-matrix diagonal reflection elements (a) and of nondiagonal reflection elements (b) for a localized case, from an ensemble of* 200 *matrices of rank* 8 *and for* $k\tau = 5$, $k = 12.4$, $\ell \simeq 50$, $L = 400$,. *The superimposed theoretical distributions, as given in ref.[10], are* $3/2\sqrt{1-x}$ *for case (a) and* $3/2(1-x)^2 {}_2F_1(3/2, 1, 3, 1-x)$ *for case (b), where* $F$ *is the hypergeometric function.*

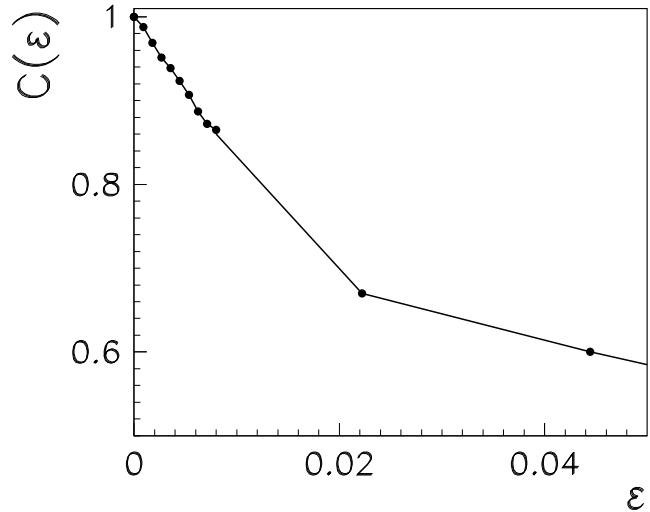

Figure 4. *Squared modulus of the normalized autocorrelation of a reflection matrix element, versus quasi-energy and for the same data as for Fig. 3.*

as a function of the quasi-energy $\epsilon$, for the same data as in Fig.3. This dependence is clearly not a Lorentzian and is suggestive of a discontinuous derivative at $\epsilon = 0$ in the limit of an infinitely long sample. We have no theoretical argument for such a behaviour. Discontinuous derivatives of the correlation curve at $\epsilon = 0$ have been reported and explained on semiclassical grounds in certain problems of chaotic scattering in the presence of classical nonexponential decay[13]. However semiclassical arguments not accounting for localization cannot work in the present case.

In conclusion, we have numerically analyzed the fluctuations of the scattering matrix in a model which displays both classical chaotic diffusion and quantum localization. This model has been shown elsewhere to exhibit some important features of transport in disordered 1-d solids; therefore the indications that emerge from our data have a bearing not only on the general problem of quantum scattering in the presence of classical chaos but also on the problem of transmission fluctuations. According to the results illustrated in this Letter, a rough qualitative agreement with the GOE model is observed in the ballistic regime but systematic deviations exist. COE statistics emerge instead in the limit of extreme localization, where the problem becomes essentially one of reflection of waves by a disordered medium. In that case, however, the correlation of $S - matrix$ elements at different quasi-energies is not a Lorentzian but it appears exhibit an angular point at the origin. It may be interesting to speculate whether this behaviour may be related to long time diffusive decay, but the presence of localization calls for great caution in that matter.